\documentclass[11pt,twoside]{article}

\usepackage{asp2006}
\usepackage{epsf}
\usepackage{lscape}
\usepackage{amssymb}
\usepackage[]{hyperref}
\hypersetup{
pdfauthor = {M. Almudena Prieto},
pdftitle = {SED of galaxy cores}
}
\usepackage[all]{hypcap}

\begin{document}
\title{The spectral energy distribution of the central parsec-scaled region of AGN}   
\author{M. A. Prieto (1), J. Reunanen (2), Th. Beckert (3),
 K. R. W. Tristram (4), N. Neumayer (4), J. A. Fernandez-Ontiveros (1),
J. A. Acosta (1) }   

\affil{(1) IAC, Tenerife;(2) Leiden Observatory; (3) MPIfR, Bonn; (4)
  MPIA, Heidelberg}

\begin{abstract} 
Genuine spectral energy distributions (SEDs) of the central few parsec
of the nearest and brightest active galaxies in the Southern
Hemisphere are presented. They are compiled from very high spatial
resolution observations in the radio (VLBA), the infrared (using
adaptive optics and interferometry) and the optical (HST).  The SEDs
are characterized by two main emission bumps peaking in the X-rays and
in the infrared respectively, as it is known from optically obscured
galactic nuclei.  Yet, the SED shape of the IR largely departs from
the one derived from large aperture data. It reveals two new features:
(1) a very sharp decay at wavelengths shortward of 2 $\mu$m, plausibly
a consequence of the heavy extinction towards the core region and (2)
a flattening in the 10-20 $\mu$m range as well as a downturn toward
longer wavelengths. Accordingly, the true bolometric luminosity of
these core regions turns out to be about an order of magnitude lower
than previously estimated on the basis of IRAS/ISO data. These
findings indicate that large aperture IR data are largely dominated by
the contribution of the host galaxy. They warn against
over-interpretations of IR/X-ray and IR/optical correlations based on
large aperture IR data which are used to differentiate AGN from normal
galaxy populations.

The new derived  IR bolometric luminosities still exceed the output energy
measured in the high energies by factors from 3 to 60. With the
expectation  that both luminosities should be comparable within an order
of magnitude, the reduced factors between both  suggests that the derived
IR luminosities are getting closer to the genuine  power output of the core.

Due to the apparent SED emission turnover in the mid-IR region, an
extrapolation of the VLBA core emission towards shorter wavelengths
closely meets the IR data. In Cen A, NGC 1068 and NGC 5506, this
extrapolation fits a power-law with an exponent of about 1/3.  This
indicates that the IR emission may not be as dust dominated as
previously thought but that it includes an important non-thermal
component.

\end{abstract}


\section{The spectral energy distributions of the nearest active galactic nuclei\label{section1}}   

High spatial resolution spectral energy distributions (SEDs) of the nearest
and brightest nuclei of galaxies accessible from the Southern
Hemisphere are presented. The selection is driven by the requirement to obtain,
in the wavelength range of 1 to 5 $\mu$m, adaptive optics observations of
the nuclei with spatial resolution scales comparable to those obtained with
radio interferometry and HST. The galaxies analyzed so far include the best
known southern AGN (e.g. Centaurus A, NGC 1068, Circinus, NGC
1097 and NGC 7582).
The compiled SEDs make use of observations with the highest spatial resolution
achievable with current instrumentation: (1) VLBA interferometry in the radio,
(2) VLT NACO adaptive optics observation in the near infrared,
(3) VLT VISIR diffraction limited observations and (4) VLTI interferometry,
both in the mid-IR, as well as (5) HST observations in the optical.
The spatial resolutions achieved in the IR have a FWHM $\lesssim $ 0.1 arcsec,
equivalent to those achieved with HST in the optical and comparable to
those in the radio. These unprecedented spatial resolutions 
in the IR allow us to pinpoint the true nucleus in each galaxy and extract its
luminosity within aperture diameters of less than 1 to 10 pc at the
distance of those targets.

\begin{figure}
\vspace{-0.5cm}
\plotone {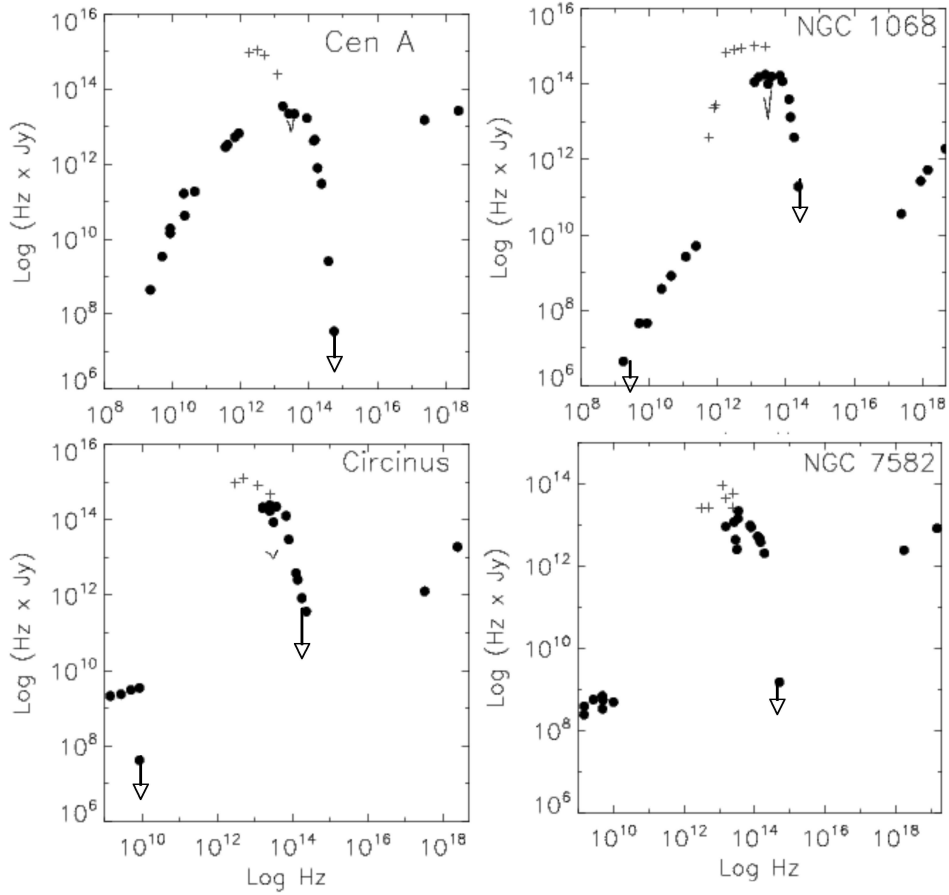}
\caption{SEDs of the central parsec-scaled region of four
  representative AGN in our study: filled points represent the highest 
  available spatial resolution data for these cores; the thin  V-shape
  line in the mid-IR region corresponds to the spectrum of the unresolved
  source measured with VLTI/MIDI (W. Jaffe et al. these proceedings); crosses refer to
  large aperture data mainly from IRAS and ISO. Further information
  is provided in Prieto, Reunanen et al. (2007, in preparation).
}
\label{figure1}
\end{figure}

Here we present SEDs that best illustrate the main result of this project,
namely the new SED shape of these galaxy cores.
The SEDs correspond to those of Cen A, Circinus, NGC 1068 and NGC 7582.
The first two galaxies are, with a distance of $\sim$ 4 Mpc, the two
nearest AGN in the Southern Hemisphere, while the latter are 4 and 7
times further away respectively.


In fig. \ref{figure1} the filled points correspond to the data with the highest
available resolution. The high energy range is covered by XMM,
BeppoSax and COMPTEL data when available. For sake of clarity, fig. \ref{figure1}
shows only frequencies up to 10 keV. The optical is covered by HST.
All these nuclei are heavily obscured, they are not detected in the optical and thus the optical data are upper limits.
They  are all, however,  unveiled in the IR. They are
unresolved at least from 2 $\mu$m on, except in the case of
Circinus \citep{P} where a resolved source of 2 pc
size is found. Fluxes from these nuclei in the 1-5 $\mu$m  range 
from  VLT/NACO adaptive optics data, in the 8 - 12 $\mu$m range from
VLTI/MIDI interferometry data (when available), and in the 10 - 20
$\mu$m range from VLT/VISIR diffraction-limited data are included
in the SEDs. The radio range is covered with VLBA/VLBI  observations
which are available for the core 
of Cen A and NGC 1068. No equivalent data exist for NGC 7582 
and Circinus. In these two cases, the available radio data are from
very large beams including both the core and  lobes of the galaxy.
As a reference for the true core emission, these large beam data are included as well in fig. \ref{figure1}.
For all cases, the figure also includes mid-IR fluxes from large aperture
IRAS and ISO data, which are marked by crosses.

\begin{figure}
\vspace{-1cm}
\plotone {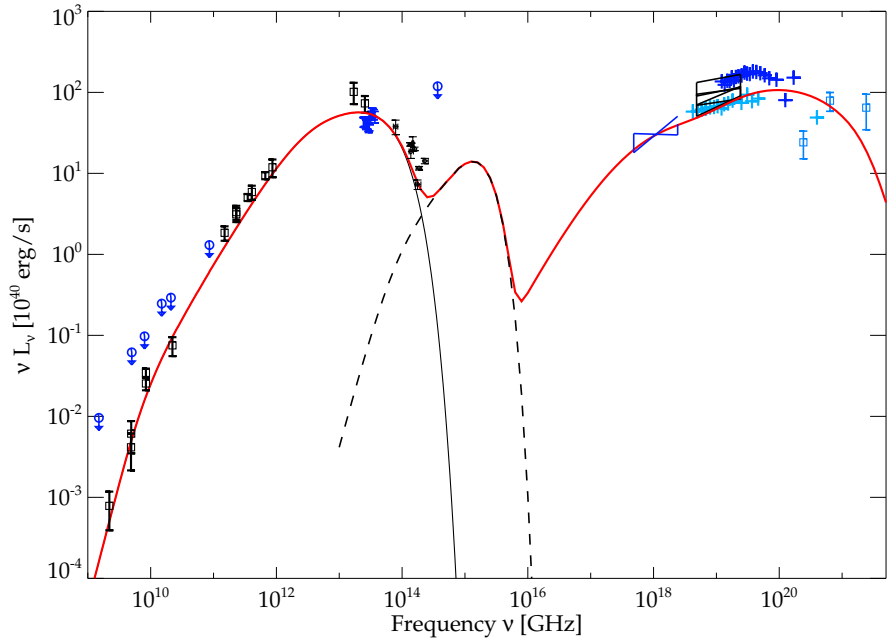}
\caption{SED of the Cen A core dereddened by $A_{\mathrm{V}}=19\,\mathrm{mag}$, this being  estimated from the silicate feature in the VLTI correlated flux
  spectrum. The continuous line is a synchrotron plus  SSC  model accounting
  for the radio and IR to the 0.1 MeV regime;  the contribution of
  a weak accretion disk at optical wavelengths, and the ``comptoionization''
  stage of the accretion disk photons in a hot corona are also
  included.  Radio points above the model are large
  aperture VLA data; millimetric data are also large aperture
  data and provided  for comparative purposes only, high energies data
  include two emission  states of Cen A. Details of the model are in 
  Beckert et al. (2007, in preparation).
}
\label{figure2}
\end{figure}

\section{Results: the new SED shape of AGN\label{section2}}
Comparing the large aperture data (crosses, fig. \ref{figure1}) with
the new high spatial resolution SEDs (filled points), two main
differences become obvious: (1) the shape of the new SED in the IR
clearly departs from that followed by the large aperture data, namely
the new SED shows a flat distribution at 2-10 $\mu$m with indications
for a fall off turnover towards lower frequencies; (2) the large aperture
data are shifted up in power by more than an order of magnitude, the
inferred bolometric luminosities are thus a factor of 2 to
10 larger than those estimated by the new high resolution data.
These two findings indicate that the mid- to far-IR luminosity
in AGN is  dominated by the host galaxy contribution and thus
IR luminosities  derived from large aperture data cannot be taken as
indicators of AGN activity or tracers of AGN populations.

\subsection{The contribution of non-thermal emission to the IR in AGN\label{subsection21}}
In some galaxies, the SED shape in the IR is such that a simple
extrapolation of the VLBI core emission towards higher  frequencies  
closely meets the IR.  Subjected to the confirmation by millimetric
data of comparable resolution, such extrapolation
indicates that the IR emission  may not be as much dust dominated as previously thought but some
other non-thermal components may be equally important. The best
example is the nucleus of Cen A: the VLBI to
IR to optical data fit a synchrotron spectrum with a spectral index
close to  0.3, similar to the case  found in the Galactic Center source $Sgr\,A^*$,
and corresponding to the index produced by  a monoenergetic
electron distribution. Accordingly, the SED of Cen A was modeled with a synchrotron
spectrum produced by a quasi monoenergetic electron distribution. Such
models have been suggested for $Sgr\,A^*$ \citep[][and references therein]{B}.
The synchrotron model (fig. \ref{figure2})
successfully accounts for the observed $\gamma$-ray emission 
which dominates the high energy SED of Cen A, due to
inverse Compton scattering of the radio synchrotron electrons.  Because of the apparent 
naked nature of the Cen A nucleus, the contribution of 
a weak  accretion disc  to account for the dereddened nuclear optical
emission is shown in fig. \ref{figure2}, however this component 
is  not  constrained with the available  data. Details of the model will be
presented in  Beckert et al. 2007, in preparation.

\end{document}